\documentclass{pasj00}
\SetRunningHead{Tsujimoto et al.}{Odd elements in the early Galaxy}

\def\Msun{M_{\odot} }
\def\cm3{{\rm ~cm}^{-3}}
\def\ltsima{$\; \buildrel < \over \sim\;$}
\def\ltsim{\lower.5ex\hbox{\ltsima}}
\def\gtsima{$\; \buildrel > \over\sim \;$}
\def\gtsim{\lower.5ex\hbox{\gtsima}}
\def\ms{$M_{\odot}$ }
\def\msp{$M_{\odot}$}

\author{Takuji \textsc{Tsujimoto}} 
\affil{National Astronomical Observatory, Mitaka-shi, 
Tokyo 181-8588, Japan}
\email{taku.tsujimoto@nao.ac.jp} 
\email{}
\author{Toshikazu \textsc{Shigeyama}}
\affil{Research Center for the Early Universe, Graduate 
School of Science, University of Tokyo, Bunkyo-ku, Tokyo 113-0033, 
Japan}
\email{shigeyama@astron.s.u-tokyo.ac.jp}
\and
\author{Yuzuru Yoshii}
\affil{Institute of
Astronomy, Graduate School of Science, University of Tokyo,
Mitaka-shi, Tokyo 181-8588, Japan}
\email{yoshii@ioa.s.u-tokyo.ac.jp}

\title{CHEMICAL EVOLUTION OF ODD ELEMENTS IN AN INHOMOGENEOUS EARLY GALAXY}

\Received{22 June 2001}
\Accepted{12 September 2001}

\begin{document}
\maketitle

\begin{abstract}

We investigate the chemical evolution of odd-numbered elements such as
sodium (Na) and aluminum (Al) during the early epochs of the Galactic
halo with the use of a model that reproduces the observed box-shaped
distribution of extremely metal-poor stars in the [Na, Al/Mg] versus
[Mg/H] plane. Our model is constructed under the assumptions that those
stars retain the elemental abundance patterns produced by {\it
individual} Type II supernovae (SNe), and that the yields of the odd
elements depend on the initial metallicity, $z$, of their SN
progenitors. As a result, recent abundance determinations that clarify
how the [Na, Al/Mg] ratios of field stars have evolved to the solar
values enable us to deduce how the yields of these odd elements depend
on $z$.  The observed trends in these abundances, in particular the very
large scatter (over 1 dex in [Al/Mg]) requires that the Al yield scales
as $m_{\rm Al}\propto z^{0.6}$ for [Mg/H]\ltsim --1.8, while the
observed [Na/Mg] trend requires that the Na yield scales as $m_{\rm
Na}\propto z^{-0.4}$ for [Mg/H]\ltsim --1.8 and $m_{\rm Na}\propto
z^{0.4}$ for [Mg/H]\gtsim --1.8. It is found that the predicted
frequency distribution of stars in the [Na/Mg] versus [Mg/H] diagram is
very sensitive to the assumed form of the primordial IMF, and that its
slope is {\it steeper than the Salpeter IMF}.  The necessity to match
the observed abundance patterns of odd elements and the frequency
distribution of extremely metal-poor stars should provide useful
constraints on nucleosynthesis calculations of metal-free massive stars
as well as on theories of their formation.
\end{abstract}

\section{INTRODUCTION}

Recent high-resolution observations of metal-poor halo stars show that
the relative abundances of certain elements of odd nuclear charge
number, specifically sodium (Na) and aluminum (Al), as compared with
that of even nuclear charge number elements like magnesium (Mg), are on
the average smaller than the solar ratio, with large star-to-star
variations.  In particular, the scatter covers a range of
$-1<$[Na/Mg]$<0$ and $-1.5<$[Al/Mg]$<0$, far exceeding errors in
abundance measurements for stars with [Mg/H]\ltsim --2
\citep{McWilliam_95} (see also Norris, Ryan, \& Beers 2001). This
observed feature of Na and Al in very metal-poor stars (which may
partially originate from abundance measurements based on very few lines,
including resonance lines) has yet to be fully explained with present
chemical evolution models of the Galactic halo.

The elements Na, Mg, and Al, in order of increasing charge number, are
produced as a result of hydrostatic carbon and neon burning in massive
stars -- supernova (SN) explosions merely expel these elements without
affecting their yields (Pagel 1997 and references therein).  While Mg is
a primary element, both Na and Al are secondary elements in the sense
that their production depends on the amount of seed heavy elements
contained in the initial chemical composition of the SN progenitor star.
Theoretical calculations of SN models indeed indicate the existence of
an odd-even effect, such that the predicted abundance ratios of [Na/Mg]
and [Al/Mg] are smaller for lower metallicity stars, where metallicity
is indicated by [Mg/H] \citep{Arnett_71, Woosley_82}.  However, in order
to make the odd-even effect compatible with the large observed variation
in [Na/Mg] and [Al/Mg] for [Mg/H]\ltsim --2, chemical inhomogeneity in
the Galactic halo gas at early epochs needs to be introduced, otherwise
[Na/Mg] and [Al/Mg] would show a very tight correlation with [Mg/H],
which is not seen.

One mechanism for introducing the required inhomogeneity is the model of
chemical evolution proposed by Tsujimoto, Shigeyama, \& Yoshii (1999,
hereafter TSY99; 2000, hereafter TSY00), in which each SN event triggers
star formation in the swept-up gas, so that newly formed stars inherit
the elemental abundance pattern of individual SNe \citep{Shigeyama_98,
Tsujimoto_98, Nakasato_00}. Argast et al.~(2000) also constructed a 3D
stochastic chemical evolution model that can incorporate the
inhomogeneity introduced by individual SN explosions.  This kind of
inhomogeneous models has successfully reproduced the observed variety of
abundance patterns of $\alpha$-elements and iron-peak elements, based on
theoretical SN yields of these heavy elements, which are presently
calculated with reasonable accuracy. It is important to note that,
because abundance patterns for extremely metal-poor stars should retain
the fossil record of the nucleosynthesis of their individual SN
progenitors, the inhomogeneous model is a useful tool for deducing the
production sites and yields of other elements for which SN models have
not yet provided consistent results (TSY00).
 
Significant progress has been made in nucleosynthesis calculations based
on SN progenitors with various metallicities and masses, but theoretical
SN yields of odd elements such as Na and Al have not yet converged to
consistency among models from different authors \citep{Woosley_95,
Limongi_98, Umeda_00}. \citet{Limongi_98} and Umeda et al.~(2000)
predict that both the synthesized masses of Al and Na for 20 \ms SN
model increase by a factor of $\sim$3--4 as the metallicity increases
from zero to $z$=0.05$z_\odot$. On the other hand, \citet{Woosley_95}
show that for a 22 \ms SN model, the amount of Al is insensitive to the
metallicity for the range of $z=0-0.01z_\odot$, whereas the amount of Na
{\it decreases} by a factor of $\sim$25 with increasing metallicity in
the same range. In addition, there has also been significant progress in
observational abundance determination of these elements, providing new
information on elemental abundance trends for both extremely metal-poor
stars as well as higher metallicity disk stars (Pilachowski, Sneden, \&
Kraft 1996; Hanson et al.~1998; Fulbright 2000). These observations show
how the [Na, Al/Mg] ratios of field halo stars have evolved to the solar
values.  The results are intriguing.  A tight correlation of [Al/Mg]
ratios with [Mg/H] starts at [Mg/H] $\sim -1.8$, with the values of
[Al/Mg] higher than those of most extremely metal-poor stars populating
the boxed-shape distribution, whereas for [Na/Mg] the trend starts to
change at [Mg/H]$\sim-1.8$ towards the solar metallicity while gradually
reducing their star-to-star variation of [Na/Mg].  This might imply a
different $z$-dependence between the Al and Na yields in the metal-poor
regime.

The purpose of this paper is to {\it empirically} determine the Al and
Na yields that can account for the observed abundance trends in the
context of the inhomogeneous chemical evolution model proposed by TSY99
and TSY00. In this model, new stars are formed following each SN event,
thus their abundance pattern is determined by the combination of heavy
elements ejected from the SN itself and those elements which are already
present in the interstellar gas swept up by the supernova remnant
(SNR). The predicted stellar abundance patterns are thus different from
those of the gas at the time when stars form.  This difference can be
quite large in the early stage of Galactic evolution, when the
metallicity in the gas is very low. The abundance ratios of
low-metallicity stars are predicted to exhibit a large star-to-star
scatter, depending in detail on the abundance patterns of SN ejecta with
different progenitor masses. In the later phase, the predicted scatter
in abundance ratios becomes smaller toward larger [Mg/H], which is
ascribed to the switch of the major contribution to stellar metallicity
from the ejecta of SNe to the metallicity in the interstellar gas swept
up by SNRs.

In \S 2 we derive the metallicity-dependent Na and Al yields. In \S 3 we
construct a model for the evolution of Na and Al that is fully
consistent with the large star-to-star variation in [Na/Mg] and [Al/Mg]
observed for metal-poor stars. Our conclusions are presented in \S 4.

\section{EMPIRICAL DETERMINATION OF SN YIELDS OF ODD ELEMENTS}

According to our understanding of the early evolution of the Galaxy,
there must presently exist long-lived second-generation stars with
surface abundance patterns that exactly reflect the yields of odd
elements from individual SNe of Population III (Pop III SNe).  Such
stars populate the branch associated with the production of element $X$
and Mg from individual Pop III SNe in the [$X$/Mg] versus [Mg/H]
diagram. We refer to this branch as the $y$-branch (the letter $y$
stands for its origination in individual SNe II yields), from which the
yield of element $X$ from Pop III SNe can adequately be derived (TSY00).

The seed element for Al production is Mg, while that for Na production
is Ne, whose abundance is unfortunately difficult to measure for
metal-poor stars. We therefore note that the odd-even effect, which
would remain only indicative in the [Na/Mg] ratio, is best investigated
with the [Al/Mg] ratio. Hence, in this section we first discuss Al and
then Na.

\subsection{Aluminum}

\begin{figure}
\begin{center}
\FigureFile(242bp,351bp){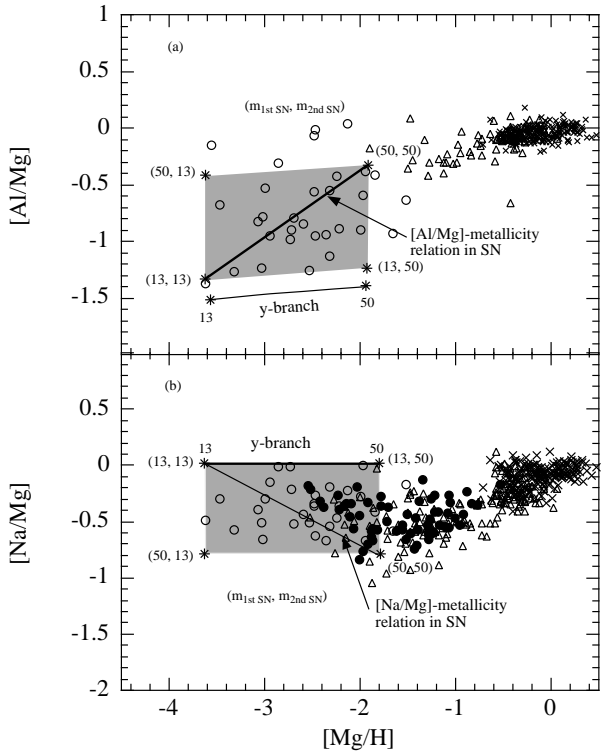}
\caption{(a) The observed correlation of [Al/Mg] with [Mg/H]
(open circles; McWilliam et al.~1995, open triangles; Fulbright 2000,
crosses; Edvardsson et al.~1993).  The thin line indicates the
$y$-branch for Pop III SNe, while the thick line represents the relation
between synthesized [Al/Mg] and {\it initial} metallicity [Mg/H] in SN
progenitors. The shaded area cornered with asterisk symbols is the
predicted location of the stars born in the first few generations (b)
The same as (a) but for Na. Other observed data (filled circles; Hanson
et al.~1998) are added.}
\end{center}
\end{figure}

Figure 1a shows the values of [Al/Mg] plotted against [Mg/H] for a
sample of stars \citep{Edvardsson_93, McWilliam_95,
Fulbright_00}\footnote{Here we limit the data to a body of tens of
stars in order to avoid the scatter originating from differences in
the values reported by different authors.} that covers the full
metallicity range. Contrary to the small scatter in [Al/Mg] for
[Mg/H]\gtsim --1.8, the values of [Al/Mg] for [Mg/H]\ltsim --1.8
scatter over 1 dex at any given [Mg/H].

There are two remarkable features clearly separated at [Mg/H]$\sim -1.8$
in the [Al/Mg] versus [Mg/H] diagram: (1) A lack of correlation between
[Al/Mg] and [Mg/H] for stars with [Mg/H]\ltsim --1.8, indicated by the
shaded area, and (2) a tight correlation for stars with [Mg/H]\gtsim
--1.8, as expected from the odd-even effect. The first feature occurs as
a result of chemical inhomogeneity in the halo (TSY99, TSY00). The
second feature occurs due to the fact that the stellar metallicity is
determined by the metallicity in the interstellar gas swept up by SNRs,
instead of that in the ejecta of individual SNe. Through the transition
from feature 1 to 2, mixing of SN products from all previous
nucleosynthesis sites with the ISM gradually weakens the chemical
inhomogeneity introduced by individual SNe. However, the gradual mixing
process is not seen in the [Al/Mg]-[Mg/H] distribution in the
metallicity range --3.6\ltsim [Mg/H]\ltsim --1.8 because the first few
generation stars overlap in this metallicity region (the shaded region
of Fig. 1a) as described below. Therefore the second feature appears for
stars with [Mg/H]\gtsim --1.8.

Theoretical SN models show that values of [Al/Mg] originating from Pop
III SNe with different progenitor masses are similar, exhibiting
differences of at most $\sim$ 0.4 dex. Thus we assume that stars born in
SNRs of Pop III progenitor stars have an almost constant [Al/Mg].  Since
there is no way to populate stars below the $y$-branch owing to the
odd-even effect, we need to locate the $y$-branch, as indicated by the
thin line, just below the data points in Figure 1a. The thick line in
this figure indicates the relation between synthesized [Al/Mg] and {\it
initial} metallicity [Mg/H], which SNe should satisfy in order to
reproduce the observed scatter in [Al/Mg]. Given this
metallicity-dependent Al yield, we can account for the resultant
distribution of long-lived stars in the [Al/Mg] versus [Mg/H] diagram.

Second-generation stars located along the $y$-branch have [Al/Mg]$\sim
-1.4$ by assumption, but have various metallicities, indicated by
[Mg/H], over a range of almost 2 dex (the thin line in Fig.~1a). For
example, because SNe with larger progenitor masses eject a larger amount
of metals, second-generation stars born in the SNR of a Pop III
progenitor star having a mass of 50 $M_\odot$ have [Mg/H]$\sim-1.8$ on
the right end of the $y$-branch, and those stars from a 13 $M_\odot$ SN
have [Mg/H]$\sim-3.6$ on the left end.

\begin{figure}
\begin{center}
\FigureFile(256bp,151bp){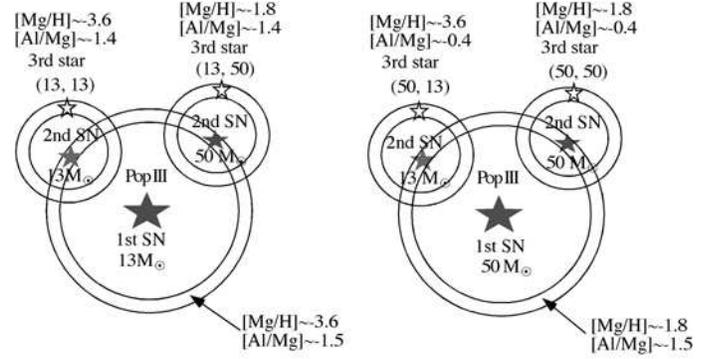}
\caption{Schematic illustration of the SN-induced star
formation model, indicating elemental abundance patterns for second- and
third-generation stars. The left shows the case for third-generation
stars born in the SNRs from progenitor stars having masses of 13 and 50
\msp, which are born in the SNR of a Pop III progenitor star having a
mass of 13 \msp. The right shows the same as the left but for a Pop III
progenitor star having a mass of 50 \msp. Concentric circles, filled
stars, and open stars denote SNR shells, SNe, and newly-formed stars,
respectively.}
\end{center}
\end{figure}

As schematically illustrated in Figure 2, third-generation stars born in
the SNR from a progenitor star with a higher metallicity, of order
[Mg/H]$\sim-1.8$, must have a higher ratio of [Al/Mg]$\sim -0.4$ due to
the metallicity-dependent Al yield. On the other hand, stars of the same
generation from a progenitor star with lower metallicity,
[Mg/H]$\sim-3.6$, must have a lower ratio of [Al/Mg]$\sim -1.4$, which
is similar to that of second-generation stars on the $y$-branch.  In
either instance, apart from [Al/Mg], stars that originated from SN
explosions must have very different metallicities of [Mg/H] if the
progenitor stars have different masses.

In this way, the location of a star in the [Al/Mg] versus [Mg/H] diagram
is determined by the combination of progenitor masses of SNe of the
preceding two generations.  In particular, third-generation stars are
distributed in the shaded area in Figure 1a, with its corners labelled
by asterisk symbols and the parenthetic quantities $(m_{\rm 1st~SN},
m_{\rm 2nd~SN})$. For example, stars near (50, 13) originate from SNe
with progenitor masses of 13 $M_\odot$, which are in turn the offspring
of Pop III SNe with progenitor masses of 50 $M_\odot$. Since stars of
later generations show similar abundance distributions, gradually
shifting towards higher [Mg/H], the first three generations suffice to
basically reproduce the observed star-to-star variation in [Al/Mg].

Consequently, the lack of a tight correlation between [Al/Mg] and [Mg/H]
for [Mg/H]\ltsim --1.8 can be understood as arising from two independent
factors: (1) The metallicity, [Mg/H], of a star, determined by the
progenitor masses of the SN from which the star was formed, and (2) the
abundance ratio [Al/Mg] of the same star, determined by the {\it
initial} metallicity of each SN progenitor.

Finally, we should note the problem of determining the Al abundance
based on resonance lines. For the metal-poor stars observed by
\citet{McWilliam_95}, the Al abundances are necessarily determined from
resonance lines, since other lines used in \citet{Fulbright_00} and
\citet{Edvardsson_93}, which give much more reliable abundances, lie
below the detection limit. Baum\"{u}ller \& Gehren (1997) have
demonstrated that the LTE analysis for these resonance lines leads to
considerable underestimation $\Delta$[Al/H]$\sim-0.65$ for stars with
[Fe/H]\ltsim$-2$. Details of how such a non-LTE effect changes the
distribution of [Al/Mg] has been investigated by Norris et al.~(2001)
and shown in their Figure 8. Their result implies that there is no
change in the overall trend of [Al/Mg] represented by the above two
features, but with a somewhat smaller scatter in [Al/Mg] for extremely
metal-poor stars, i.e., $\Delta$ [Al/Mg]$\sim$0.7 rather than $\Delta$
[Al/Mg]$\sim$1.0.

\subsection{Sodium}

Figure 1b shows the values of [Na/Mg] plotted against [Mg/H] for
metal-poor halo stars \citep{McWilliam_95} and disk stars
\citep{Edvardsson_93}. Thanks to the available [Na/Mg] data for
--2.5\ltsim[Mg/H]\ltsim --1 (Hanson et al.~1998, which is a revised
version of data from Pilachowski et al.~1996) for halo and disk stars,
as well as the data of \citet{Fulbright_00}, we clearly see the
observational trend of [Na/Mg] over the whole metallicity range. We
stress that the abundance determinations of Na by the above authors,
which determine the overall trend of [Na/Mg], are based on lines that
are little affected by conditions of non-LTE.  A smaller scatter in
[Na/Mg] compared with [Al/Mg] for [Mg/H]\ltsim --1.8 indicates that the
Na yield is less sensitive to metallicity than the Al yield in the
metal-poor regime.

The box-shaped distribution for [Mg/H] \ltsim --1.8 can in principle be
reproduced by either the horizontal $y$-branch on the bottom side of the
box, and the upward diagonal reflecting larger Na yields for higher
metallicity (similar to the Al case in Fig.~1a), or the horizontal
$y$-branch on the top side of the box and the downward diagonal
reflecting a smaller Na yield for higher metallicity stars.  However,
the requirement of a continuous transition at [Mg/H] $\sim$ --1.8 (e.g.,
from feature 1 to feature 2 --see \S 2.1--) results in our preference of
the latter alternative.  The adopted $y$-branch is shown by the thin
line in Figure 1b, and the required metallicity-dependence of the Na
yield is indicated by the thick line, which seems to be at variance with
expectations from the odd-even effect.

We note that current SN models give discrepant results for the Na yield;
Umeda et al. (2000) claimed an increasing trend of Na yield as a
function of metallicity, but Woosley \& Weaver (1995) claimed an
decreasing trend in the metal-poor regime followed by an increasing
trend towards higher metallicity (see Fig.18 of Timmes, Woosley, \&
Weaver 1995). Although this theoretical discrepancy needs to be resolved
by more detailed nucleosynthesis calculations, the observed [Na/Mg]
distribution for metal-poor stars might provide compelling evidence in
favor of the decreasing trend of Na yields.

\section{INHOMOGENEOUS CHEMICAL EVOLUTION OF ODD ELEMENTS IN 
THE GALACTIC HALO}

In this section we discuss the inhomogeneous chemical evolution of odd
elements in the Galactic halo, based on the formulation presented in
TSY99. The essence of the picture is that the star-forming process is
confined in separate clouds which make up the entire halo, and that the
chemical evolution in these clouds proceeds through repetition of
successive sequence of SN explosion, shell formation, and star formation 
therein. All Pop II stars are assumed to form in SNR shells. Heavy elements
ejected from an SN are assumed to be trapped and well-mixed within the
SNR shell. Some of these elements go into stars of the next generation,
and the rest is left in the gas that will be mixed with the ambient
medium. Thus the abundance of heavy element $i$ in stars,
$z_{i,\ast}(m,\, t)$, born at time $t$ from an SNR shell with
progenitor mass $m$, is defined as
\begin{equation}
\label{eqn:massms}
z_{i,\ast} ({m{\rm ,}\, t})={M_{z_i}(m)+z_{i, \rm g}(t)M_{\rm
sw}({m{\rm ,}\, t}) \over M_{\rm ej}(m)+M_{\rm sw}({m{\rm ,}\, t})},
\end{equation} 
where $M_{z_i}(m)$: the mass of synthesized heavy element $i$
ejected from a star with mass $m$, $z_{i, \rm g}(t)$: the abundance of
heavy element $i$ in the gas, $M_{\rm sw}(m,t)$: the mass of the shell
formed at time $t$ from an SN with progenitor mass $m$, and $M_{\rm
ej}(m)$: the mass of the SN ejecta. Details can be found in TSY99.

\begin{figure}
\begin{center}
\FigureFile(235bp,187bp){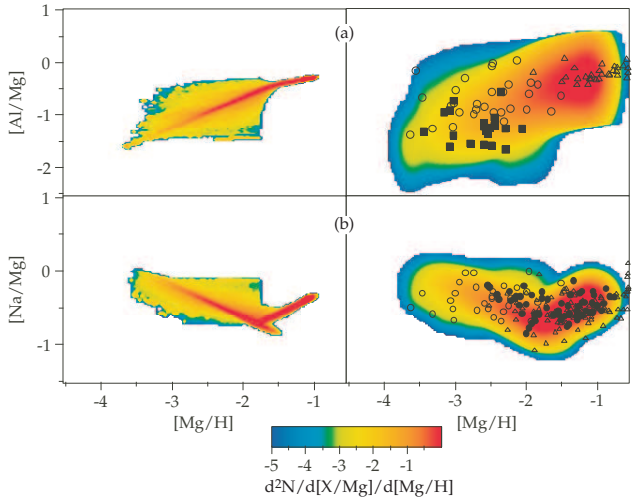}
\caption{(a) Color-coded frequency distribution of the
long-lived stars in the [Al/Mg]-[Mg/H] plane, convolved with a Gaussian
having $\sigma=0.3$ dex for [Al/Mg] and $\sigma=0.1$ dex for [Mg/H]
(right panel), and without convolution (left panel). The symbols
represent the same data which as in Figure 1. Other observed data
(filled squares; Ryan, Norris, \& Beers 1996) are added.  (b) The same
as (a) but for Na. The [Na/Mg] ratios are convolved with $\sigma=0.1$
dex in the right panel.}
\end{center}
\end{figure}

The free parameters in the inhomogeneous model are the mass fraction
$x_{\rm III}$ of metal-free Pop III stars initially formed in each
cloud, and the mass fraction $\epsilon$ of stars formed in the dense
shells swept up by each SNR.  We take $x_{\rm III}=2.5\times10^{-4}$,
which is consistent with the observed level of [Ba/Mg] for the
$i$-branch (TSY00).  We take $\epsilon=4.3\times10^{-3}$ in order to
reproduce the observed metallicity distribution function of metal-poor
halo stars (TSY99). The initial stellar mass function (IMF) used here is
a Salpeter one with upper and lower mass limits of $50\Msun$ and
$0.05\Msun$, respectively. The lower mass limit of stars that explode as
SNe is taken to be $10\Msun$.

First, relating the synthesized Mg mass $m_{\rm Mg}$ to the mass $m$ of
the SN progenitor star, and also to the Mg abundance of a SNR shell from
which stars are born \citep{Shigeyama_98}, the $y$-branch in the [Al/Mg]
versus [Mg/H] diagram can be converted into the Al yield of Pop III SNe,
denoted by $m_{\rm Al,III}$ as a function of $m$.  Next, using the
diagonal shown by the thick line in Figure 1a, the metallicity-dependent
Al yield is taken as $m_{\rm Al}(m, {\rm [Mg/H]})=m_{\rm
Al,III}(m)\times 10^{0.6({\rm [Mg/H]}+3.7)}$, or $m_{\rm Al}\propto
z^{0.6}$, scaling with some power of the initial metallicity $z$ of the
SN progenitor star. If the observed scatter of [Al/Mg] for metal-poor
stars is reduced as expected by the non-LTE effect on the Al abundances
(Norris et al.~2001), the metallicity-dependence on the Al yield becomes
small. For instance, the [Al/Mg] distribution obtained by Norris et
al.~(2001), which is taken into account non-LTE effect by Baum\"{u}ller
\& Gehren (1997), yields $m_{\rm Al}\propto z^{0.4}$.

Adopting the Al yield $m_{\rm Al}$ as a function of $m$ and $z$ obtained
as above, we calculate the expected frequency distribution of stars in
the [Al/Mg] versus [Mg/H] diagram. The left panel of Figure 3a shows the
color-coded distribution after normalization to unity when integrated
over the entire area.  In order to enable a direct comparison with the
data, the distribution has to be convolved with Gaussian errors having
$\sigma=0.3$ dex for [Al/Mg] and $\sigma=0.1$ dex for [Mg/H]. The right
panel of Figure 3a shows this convolved distribution, which agrees well
with the distribution of the observed data.

Similar to the case of Al, by using the $y$-branch and the diagonal
metallicity trend for Na in Figure 1b, we obtain the
metallicity-dependent Na yield as $m_{\rm Na}\propto z^{-0.4}$ for
[Mg/H]\ltsim --1.8, which reverses to become $m_{\rm Na}\propto z^{0.4}$
for [Mg/H]\gtsim --1.8. We then calculate the expected frequency
distribution of stars to be compared with the data in the [Na/Mg] versus
[Mg/H] diagram, and find that a steeper IMF for metal-free Pop III stars
enhances the variation in [Na/Mg]. As discussed in \S 2.1, the stellar
distribution in the [Na/Mg]-[Mg/H] plane for [Mg/H]$\leq$ --1.8 can be
basically described by third-generation stars. All second-generation
stars populate on the thick line [Na/Mg]$\sim0$ in Figure 1b. The
[Na/Mg] ratio of each star of the third generation is determined by the
yields of these elements from the preceding SN (2nd SN in Fig.~1b) and
the amounts of Na and Mg in the ISM eventually swept up by this SN. It
is found that the Salpeter IMF makes the contribution of Na and Mg in
the ISM to the [Na/Mg] ratios of third-generation stars significant, and
the values of [Na/Mg] approach $\sim0$ (thick line in Fig.~1b).  On the
other hand, a steep IMF decreases the fraction of Pop III stars that
undergo SN explosion, so that Na and Mg in the ISM cannot contribute to
the [Na/Mg] ratios of third-generation stars. As a consequence,
third-generation stars could populate the full range of the shaded
region in Figure 1b. Therefore, this dependence can be used to constrain
the IMF slope for the massive part, i.e., $m\geq 10$\ms from the [Na/Mg]
versus [Mg/H] diagram. For example, the IMF slope index for PopIII stars
of $x=-1.8$ results in the frequency distribution shown in Figure 3b. To
be consistent with the box-shaped distribution observed for [Mg/H]\ltsim
--1.8, we conclude that the primordial IMF should be more inclined than
the Salpeter IMF.

\section{SUMMARY AND DISCUSSION}

Based on the hypothesis that early generations of stars are born from
individual SNRs in the young Galaxy, we have explored the nature of the
abundance trends seen in odd elements such as Na and Al. We have found
that the combination of SN-induced star formation and
metallicity-dependent yields of odd elements from SNe can account for
the large variation in [Na/Mg] and [Al/Mg] observed for extremely
metal-poor halo stars.  The Al yield, required to reproduce the observed
star-to-star variation spanning $-1.5<$[Al/Mg]$<0$ for [Mg/H]\ltsim
--1.8, should scale with the initial metallicity $z$ of SN progenitor
stars as $m_{\rm Al}\propto z^{0.6}$.  The Na yield, required to
reproduce the variation of $-1<$[Na/Mg]$<0$, should scale as $m_{\rm
Na}\propto z^{-0.4}$ for [Mg/H]\ltsim --1.8 and $m_{\rm Na}\propto
z^{0.4}$ for [Mg/H]\gtsim --1.8.
 
An inverse $z$-trend of the Na yield, which decreases with increasing
$z$ for the metal-poor stars, is obtained from observations of [Na/Mg]
-- this may be understood if the Na production on its seed Ne is
exceeded by the $\alpha$-capture of Ne into Mg. Our result agrees with
nucleosynthesis calculations of massive stars by \citet{Woosley_95}, but
it is, on the other hand, at variance with other calculations by Umeda
et al.~(2000). A definitive confirmation of our result therefore
requires more detailed nucleosynthesis calculations.

We found that the box-shaped distribution of extremely metal-poor stars
in the [Na/Mg] versus [Mg/H] diagram constrains the slope of primordial
IMF to be steeper than the Salpeter IMF. This suggests that metal-free
stars beyond $50\Msun$ are significantly deficient, as advocated from
theoretical study of star formation in metal-free gas \citep{Yoshii_79,
Yoshii_86}.

[Al/Mg] ratios that are above the solar value are not predicted in our
model of inhomogeneous chemical evolution, unless a large
observational error is allowed for.  However, Iwamoto et al.~(2001)
showed that if efficient internal mixing processes occur, the Al
abundance of a stellar surface whose metallicity is [Mg/H]$\sim -3.5$
is enhanced up to [Al/Mg]$\sim 0$ during later phases of stellar
evolution.  Their result implies the possible existence of stars
having more enhanced Al abundances if stronger mixing takes place in
their stellar interiors.  In this regard, it should be noted that the
red giant stars of globular cluster M13 exhibit an anomalously high
[Al/Mg] ratio in the abundance range 0\ltsim [Al/Mg]\ltsim 1.3
(e.g., Kraft et al.~1997; Norris \& Da Costa 1995), which does not
overlap at all with $-1.5$\ltsim [Al/Mg]\ltsim 0 observed for field
halo stars.  Moreover, such an Al overabundance relative to Mg in
globular cluster stars is anti-correlated with metallicity
\citep{Gratton_00}, which indicates the enhanced production of odd
elements by $p$-capture processes taking place deep inside
\citep{Cavallo_96}, followed by mixing due to meridional circulation
and turbulent diffusion driven by stellar rotation \citep[Zahn's
mechanism]{Denissenkov_00}.  Some indication that globular cluster
stars indeed rotate while field stars do not \citep{Kraft_97} partly
justifies preferential acquisition of angular momentum by protostellar
fragments in dense environments. Thus, the origin of anomalously high
[Al/Mg] ratios seen in globular cluster stars is more local and can be
separated from that seen in field halo stars related to the global
chemical evolution of the Galactic halo.

\bigskip

We are grateful to Timothy C. Beers for his careful reading and
helpful comments. This work has been partly supported by COE research
(07CE2002) and a Grant-in-Aid for Scientific Research (11640229) of
the Ministry of Education, Science, Culture, and Sports in Japan.

\end{document}